\documentclass[twocolumn]{aastex631}

\newcommand\rprime{r$^{\prime}$}
\newcommand{\imageinfo}{North is up; east is to the left; 10$^4$ km scalebars are in the lower left; midpoints of exposure times in UT hours are at bottom right.  These images are color enhanced to show contrast.}
\newcommand\cnmass{4.4} %kg
\newcommand\ctwomass{4.6--4.8}
\newcommand\totalmass{9.2}

\graphicspath{{./}{figures/}}

\shorttitle{Repeating CN feature from 45P}
\shortauthors{Springmann et al.}

\begin{document}

\title{Repeating gas ejection events from comet 45P/Honda-Mrkos-Pajdu\v{s}\'{a}kov\'{a}}

\correspondingauthor{Alessondra Springmann}
\email{sondy@lpl.arizona.edu}

\author[0000-0001-6401-0126]{Alessondra Springmann}
\affiliation{Lunar \& Planetary Laboratory, University of Arizona
1629 E University Blvd., 
Tucson, AZ 85721-0092, USA}

\author{Walter M.~Harris}
\affiliation{Lunar \& Planetary Laboratory, University of Arizona
1629 E University Blvd., 
Tucson, AZ 85721-0092, USA}

\author{Erin L.~Ryan}
\affiliation{SETI Institute,
189 Bernardo Ave.,
Mountain View, CA 94043-5138, USA}

\author{Cassandra Lejoly}
\affiliation{Lunar \& Planetary Laboratory, University of Arizona
1629 E University Blvd., 
Tucson, AZ 85721-0092, USA}

\author{Ellen S.~Howell}
\affiliation{Lunar \& Planetary Laboratory, University of Arizona
1629 E University Blvd., 
Tucson, AZ 85721-0092, USA}

\author{Beatrice E.A.~Mueller}
\affiliation{Planetary Science Institute,
1700 East Fort Lowell Rd., 
Tucson, AZ 85719-2395 USA} 

\author{Nalin H.~Samarasinha}
\affiliation{Planetary Science Institute, 
1700 East Fort Lowell Rd.,
Tucson, AZ 85719-2395 USA} 

\author{Laura M.~Woodney}
\affiliation{California State University, San Bernardino,
5500 University Pkwy., 
San Bernardino, CA 92407-2397, USA}

\author{Jordan K. Steckloff}
\affiliation{Planetary Science Institute,
1700 East Fort Lowell Rd.,
Tucson, AZ 85719-2395 USA} 

\begin{abstract}
Studying materials released from Jupiter-family comets (JFCs) --- as seen in their inner com\ae, the envelope of gas and dust that forms as the comet approaches the Sun --- improves the understanding of their origin and evolutionary history. As part of a coordinated, multi-wavelength observing campaign, we observed comet 45P/Honda-Mrkos-Pajdu\v{s}\'{a}kov\'{a} during its close approach to Earth in February 2017. Narrowband observations were taken using the Bok 90'' telescope at Kitt Peak National Observatory on February 16 and 17 UT, revealing gas and dust structures. We observed different jet directions for different volatile species, implying source region heterogeneity, consistent with other ground-based and \textit{in situ} observations of other comet nuclei. A repeating feature visible in CN and C$_2$ images on February 16 was also observed on February 17 with an interval of $7.6\pm0.1$ hours, consistent with the rotation period of the comet derived from Arecibo Observatory radar observations.  The repeating feature's projected gas velocity away from the nucleus is 0.8 km s$^{-1}$, with an expansion velocity of 0.5 km s$^{-1}$. A bright compact spot adjacent to the nucleus provides a lower limit of the amount of material released in one cycle of $\sim$\totalmass{} kg, depending on composition --- a quantity small enough to be produced by repeated exposure of nucleus ices to sunlight. This repeating CN jet, forming within 400 km of the nucleus, may be typical of inner coma behavior in JFCs; however, similar features could be obscured by other processes and daughter product species when viewed from distances further than the scale length of CN molecules.
\end{abstract}

\keywords{comets --- observations --- jets}

%%%% INTRODUCTION %%%%
\section{Introduction} \label{sec:intro}
As comets approach the Sun, gases sublimate from ices and form a gas-rich coma.  Processes occurring in the inner coma provide information about volatile production from the nucleus, as well as information about photochemical evolution of short-lived species in the coma. 
We define the \textit{inner coma} as the region where collisions between gas molecules occur, affecting the velocities of other particles.  Outside of the inner coma, particle velocities are not affected by collisions.
Inner-coma processes occur quickly and/or on small spatial scales and generally can neither be temporally nor spatially resolved with ground-based telescopes. These processes are often blocked from view by other photodissociation product species in the outer coma.  
Short of \textit{in situ} exploration, our best opportunity to study inner coma processes is for a comet to experience a close approach to Earth, as outer coma daughter species obscure short-lived inner coma species.
 
Jupiter-family comet 45P/Honda-Mrkos-Pajdu\v{s\'{a}kov\'{a} (45P hereafter) passed within 0.14 au of Earth between 2016 and 2017, providing excellent circumstances for observing its inner coma with ground-based telescopes.}
Jupiter-family comets (JFCs) have orbital periods less than 20 years; they originate as trans-Neptunian objects that evolve into Centaurs and eventually migrate into the inner Solar System \citep{Duncan1997,Duncan2004,Dones2015,Tiscareno2003,DiSisto2007,Sarid2019,Steckloff2020}.  Because of their trans-Neptunian origin, studying JFCs can provide insight into Solar System evolutionary processes.

Visible-wavelength observations of 45P during its close approach were conducted as part of a multi-wavelength, global, coordinated observing campaign \citep{Harris2017}.  Observing 45P in 2016--2017 was one of the last opportunities for close study of inner coma processes in JFCs for the next several decades.  
The goals of the campaign were to use ground-based observations in visible, infrared, radar, and radio wavelengths to obtain holistic, multi-wavelength understandings of the behavior of inner coma gas and dust species and their interactions with the comet nucleus.

\subsection{Comet 45P Overview}

Comet 45P is approximately 0.6--0.65 km in diameter, with a 7.6 $\pm$ 0.5 h rotation period \citep[][]{CBET2017}. It has low H$_2$O production rates compared to similar JFCs and is depleted with respect to H$_2$O in some volatile species, including CN and H$_2$O \citep{Fink2009}.
 
Other observations taken during the 2017 apparition of 45P on January 6--8, shortly after perihelion on 2017 December 31 show volatile species preferentially released from one section of the nucleus and processing of nucleus ices before being incorporated into the broader coma \citep{DiSanti2017}.  \citet{Moulane2018}, observing from 2017 February 10--March 30,  confirmed low gas and dust activity and did not detect any outbursts; and that 45P, like 41P, has a carbon-bearing species composition typical of comets.  
Observations by \citet{DelloRusso2020}, taken 2017 February 13 and 19, indicated ``a relatively symmetric and uniform coma [...]~with small spatial differences noted between some volatile species.'' Observations from previous as well as the 2017 apparitions measuring C$_2$ and CN production rates, as well as C$_2$H$_2$ and HCN abundances, appear consistent with HCN being the primary parent of CN.
\citet{DelloRusso2020} noted the dust-to-gas ratio increased from February 13 to 19.

In this paper we present observations in filters targeting volatile species and dust of comet 45P.  In Section \ref{sec:observations} we provide an overview of CN, C$_2$, OH, and dust observations taken at Kitt Peak National Observatory.  Specifics of the standard reduction and image analysis techniques used are reported in Section \ref{sec:reduction}.  Section \ref{sec:results} describes a repeating, discontinuous CN feature, as well as features seen in other wavelengths.  We discuss in Section \ref{sec:discussion} possible causes of the repeating feature, radar observations of 45P, and comparison of these 45P observations to those taken of other comets.  We summarize the results from this campaign and for observations of future close approaches of JFCs to Earth in Section \ref{sec:conclusions}.

%%%% OBSERVATIONS %%%%
\section{Observations} \label{sec:observations}

Observations were conducted using the Bok 90''/2.3-meter telescope at Kitt Peak National Observatory outside of Sells, Arizona, USA, on 2017 February 16 and 17 UT (Table \ref{tab:circumstances}).
Observations were also obtained at the Kuiper Telescope, Large Binocular Telescope, MMT Observatory, Arecibo Observatory, and Arizona Radio Observatory as part of a coordinated, multiwavelength campaign to characterize three Jupiter-family comets \citep{Harris2017}.
Comet 45P was observed 1.5 months after perihelion and 5--6 days after perigee to identify any changes in the rotation of the nucleus and coma evolution \citep[e.g.][]{Howell2018,Schleicher2019}.

Narrowband filters (CN, C$_2$, OH, and blue continuum (BC)) originally designed for the comet Hale--Bopp (C/1995 O1) observational campaign \citep{Farnham2000} were used to ensure consistency of data across observing sites and telescopes. Both BC as well as broadband SDSS-\rprime{} were used as a proxy for dust.  
The BC images show the same features as the \rprime{} images, however, the \rprime{} exposures have more temporal coverage and better signal-to-noise ratios with shorter exposure times (and thus shorter star trails) with \rprime{} than BC images.  We limit the analysis presented here to \rprime{}-filter images, as the images are otherwise indistinguishable.
Although \rprime{} is a broadband filter and has some gas contamination, unless the comet is very dust-poor, the flux in \rprime{} images is dominated by scattered dust, not by gas.  
Using \rprime{} as a proxy for dust is a reliable method for comets with a moderate to high dust-to-gas ratio \citep{Farnham2000}, such as 45P \citep{Fink2009}, and has been successfully used for other comets \citep[e.g.][]{Mueller2013,Knight2021}.

Over the course of two nights, 19 science images were taken: 16 on February 16, and only three on February 17 due to weather.  The close approach of comet 45P to Earth allowed for plate scales of $\sim$40 km/pixel, or resolutions of $\sim$80 km/arcsecond.

\begin{footnotesize}
\begin{table*}[ht]
\centering 
    \caption{Observing circumstances for comet 45P on UT 16 and 17 February 2017 from the 90" Bok telescope on Kitt Peak, Arizona, USA.} 
        \centering
        \begin{tabular}{l c c c c c c c}
            \hline
            Date   & UT           & Days from  & Days from & $R$  & $\Delta$ & Elong.       & Phase \\
            ~      & Start--End   & perihelion & perigee   & (au) & (au)     & ($^{\circ}$) & ($^{\circ}$) \\
            16 Feb & 07:04--12:35 & 47          & 5 & 1.05 & 0.11     & 124--126          & 50 \\
            17 Feb & 07:55--08:20 & 48          & 6 & 1.07 & 0.11     & 130--131          & 44 \\
            % \tnote{a}
            \hline
        \end{tabular}
        \label{tab:circumstances}
\end{table*}
\end{footnotesize}

\section{Image Reduction \& Analysis} \label{sec:reduction}
Standard techniques were followed for bias subtraction and flat fielding using dome flats.
Measured efficiencies of the filters and the color gradient from r$^{\prime}$ to BC were used to determine appropriate subtraction scaling.  Dust contamination is an issue with C$_2$ images. To remove dust contamination, scaled r$^{\prime}$ images taken as close as possible in time to the C$_2$ images were photometrically scaled using methods outlined in \citet{Farnham2000} and subtracted from C$_2$ images.  Both CN and OH filters have a minimal, low-contrast contribution from dust: this does not affect the analysis of the images presented here. CN integration times were 900--1200 s; BC, 1200 s; C$_2$, 1200; and r$^{\prime}$, 180. The CN image exposure times are different as a resulting of optimizing the signal-to-noise ratio of the images.

Images obtained on the first night were taken under photometric conditions.  Observations of standard star HD 30246 was used for calibrating all comet images.  Sky background values were separated from the coma background by extending a radial profile until it becomes indistinguishable from flat sky at the signal-to-noise ratio of the background levels.  This may result in slight excess of background signal subtraction, though this amount is less than the variation in sky flat signal (1--2\%).  Due to the large area of the detector, we were able to measure background signal levels sufficiently far away from the comet to eliminate coma contamination.

\begin{figure*}
\begin{center}
\includegraphics[width=2.1\columnwidth]{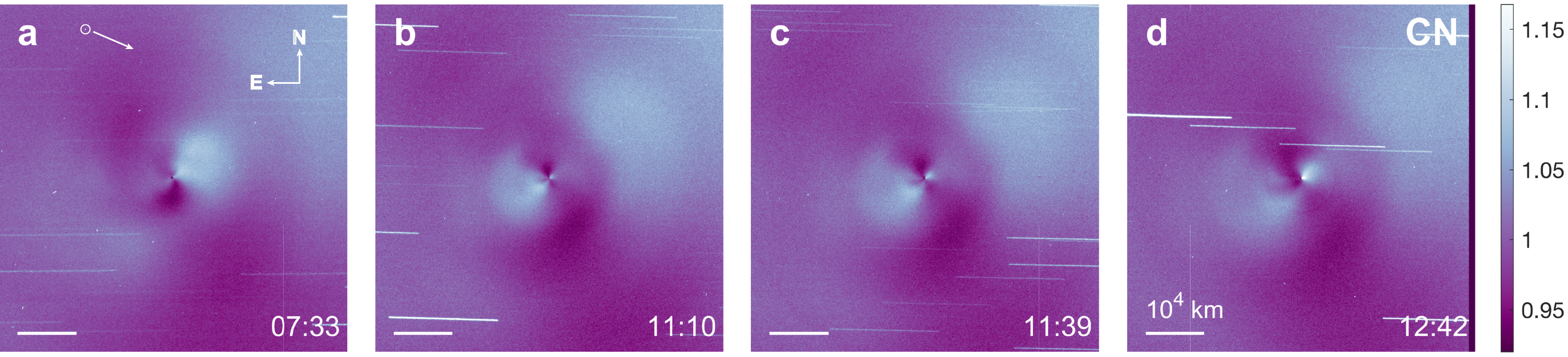}
\caption{Four enhanced images of 45P taken in the CN filter on 2017 February 16 UT. Integration times are 900 s for (a) and 1200 s for (b) to (d).  \imageinfo\ Different CN exposure integration times were used to optimize the signal-to-noise ratios of the images.  The Sun position angle of 262$^{\circ}$ is shown in the upper left of (a).  %The southeast feature (lower left) quadrant of the images and the repeating northwest feature (upper right) are visible.  
Star trails are visible as bright, straight, parallel lines.  The images are color-enhanced to show contrast, with purple corresponding to lower fluxes and light blue corresponding to higher fluxes.}
\label{fig:cnprogression}
\end{center}
\end{figure*}

Inner coma structures are only a few percent above background coma signal levels, so image enhancement techniques are required for monitoring evolution of these structures, determining the periodicity of features, and measuring outflow velocities of species.  
Because the original, reduced science images are dominated by background comet flux for the species concerned, we used radial profile-based enhancement methods described in \citet{Samarasinha2014}.  %: %Specifically, science images were divided by the median image to enhance any potential coma features.   
Sky background values were separated from the coma background by creating a median pixel value radial profile centered on the nucleus until it became indistinguishable from flat sky at the signal-to-noise ratio of the background levels.  Science images were then divided by the median image radial profile to enhance coma features (Figure \ref{fig:cnprogression}).  This may result in slightly too much background signal subtraction, but by less than the variation in sky flat signal (1--2\%).  Owing to the large area of the detector, we were able to measure background signal levels sufficiently far away from the comet to eliminate coma contamination.

%%%%%%%%%%%%%%%%%%%%%%%%%%%%%%%%%%%%%
%%%%%%%%%%%%%% RESULTS %%%%%%%%%%%%%%
%%%%%%%%%%%%%%%%%%%%%%%%%%%%%%%%%%%%%

\section{Results} 
\label{sec:results}

Preliminary inspection of the raw, un-enhanced images revealed a bright compact spot adjacent to the nucleus (Section \ref{sec:brightcompact}). 
Enhanced images show a repeating feature in CN associated with the bright compact spot not seen in other filters (Section \ref{sec:repeating}).
C$_2$ and dust images show jets in the same direction as the repeating CN feature, but there is no detection of repeating features (Sections \ref{sec:otherwavelengths}; \ref{sec:dust}).  

\subsection{A bright compact release of material}
\label{sec:brightcompact}

\begin{figure*}
\begin{center}
\includegraphics[width=2.1\columnwidth]{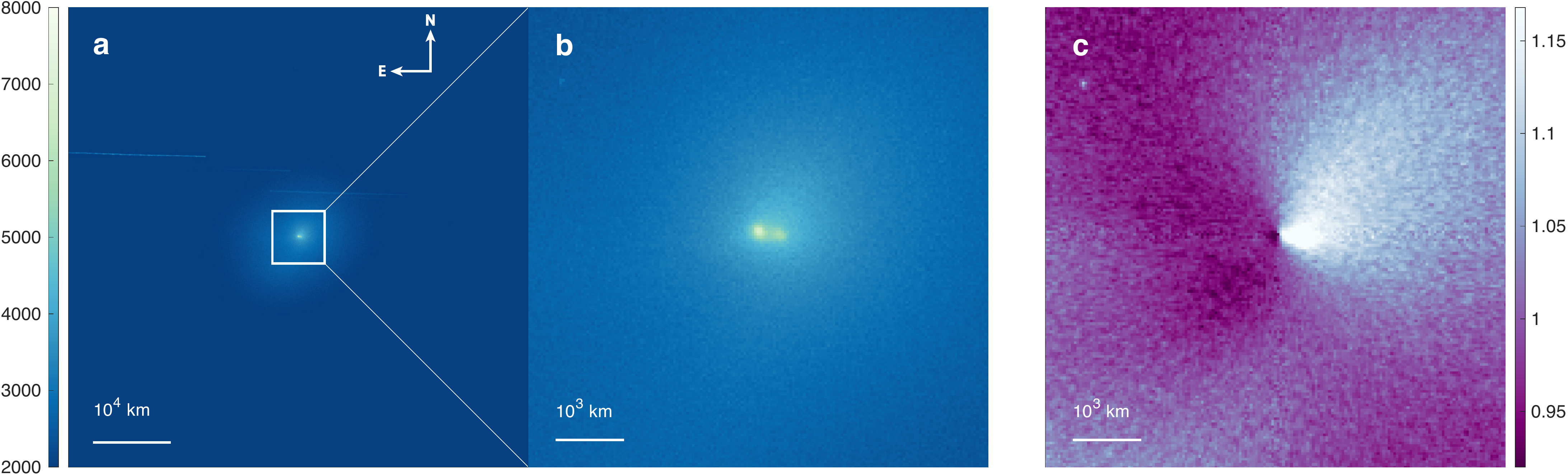}
\caption{The final 900-s CN exposure of February 16 with a exposure midpoint of 12:42 UTC. The first two images (a, b) are reduced but un-enhanced. The final image (c) has been enhanced by azimuthal subtraction.  The bright nucleus is visible (a), and closer inspection (b) reveals a second bright spot detached and to the west of the nucleus.  The enhanced image (c) shows the detached bright spot signal as part of a larger bright, discontinuous feature moving away from the nucleus toward the northwest.}
\label{fig:raw}
\end{center}
\end{figure*}

A bright compact spot is visible in an un-enhanced image (Figure \ref{fig:raw}).  Preliminary inspection of the raw, un-enhanced images revealed a bright compact spot with a projected area visible above the background coma of $7.8 \times 10^{4}$ km$^2$ immediately to the west of the nucleus in the final CN image of the night before dawn taken on February 16 12:42 UTC (Figure \ref{fig:raw}).  
% Area corresponds to circle with a radius of 4 pix
The area of the bright compact spot was measured by counting pixels with signal above the background in the reduced, un-enhanced image.  Measuring signal obscured by the coma would affect photometric accuracy; thus, this measurement is a lower bound on  the area of the bright compact spot.

Possible explanations of the bright compact spot include jitter, optical ghosts, and stars or other background objects: all of these would have resulted in trailing visible in the image.  Careful alignment of the nucleus during image enhancement reveals a halo of emission around the bright compact spot separate from the nucleus position (Figure \ref{fig:raw}), implying cometary-like behavior of the bright compact spot, which would be unlikely if it were a background asteroid or other astronomical object.  It is likely that this bright compact spot would have been visible in a C$_2$ image; however, our observing strategy close to sunrise prioritized CN and \rprime\ images.

We calculated the production rate of material in the bright compact spot (Figure \ref{fig:raw}) as approximately $3.7 \times 10^{24}$ molecules s$^{-1}$, consistent with the overall flux for the inner coma of $2.2 \times 10^{24}$ molecules per second measured by \citet{Moulane2018}, using fluorescent efficiencies for CN at $R = 1.05$ au \citep{Schleicher2010}.\footnote{A calculator for comet fluorescence efficiencies/g-factors for common comet volatile species based on \citet{Schleicher2010} is available at: \url{https://asteroid.lowell.edu/comet/cover_gfactor.html}}  Our calculated production rate is likely an underestimate because we only measured signal detectable above background levels.  

Using the production rate of the bright compact feature we calculated \cnmass{} kg as a lower limit of the total CN mass ejected from the nucleus. 

\subsection{Repeating, Discontinuous CN Feature}\label{sec:repeating}

Image enhancement revealed that the bright compact spot is part of a larger CN cloud, roughly elliptical in shape, retreating from the nucleus (Figure \ref{fig:cnprogression}).  We see CN material brightening at the nucleus then retreating as a large cloud of material originating from the nucleus.  The cloud brightness ramps up rapidly, and the material moving away from the nucleus has a discontinuity in its morphology: the material does not move as one continuous jet emission from the nucleus along the northwest direction.  A plume of material is initially visible in the first images taken on February 16 (Figure \ref{fig:cnprogression}a), which eventually detaches from the nucleus as a 24,000 km $\times$ 16,000 km bean-shaped cloud with the short axis aligned toward the nucleus (Figure \ref{fig:cnprogression}b).  A second plume forms (Figure \ref{fig:cnprogression}c), visible as a bright mass of CN gas detaches from the nucleus (Figures \ref{fig:cnprogression}d, \ref{fig:raw}).  Ultimately, three separate clouds of CN gas are visible moving away from the nucleus.  
  
\begin{figure*}
\begin{center}
\includegraphics[width=2.1\columnwidth]{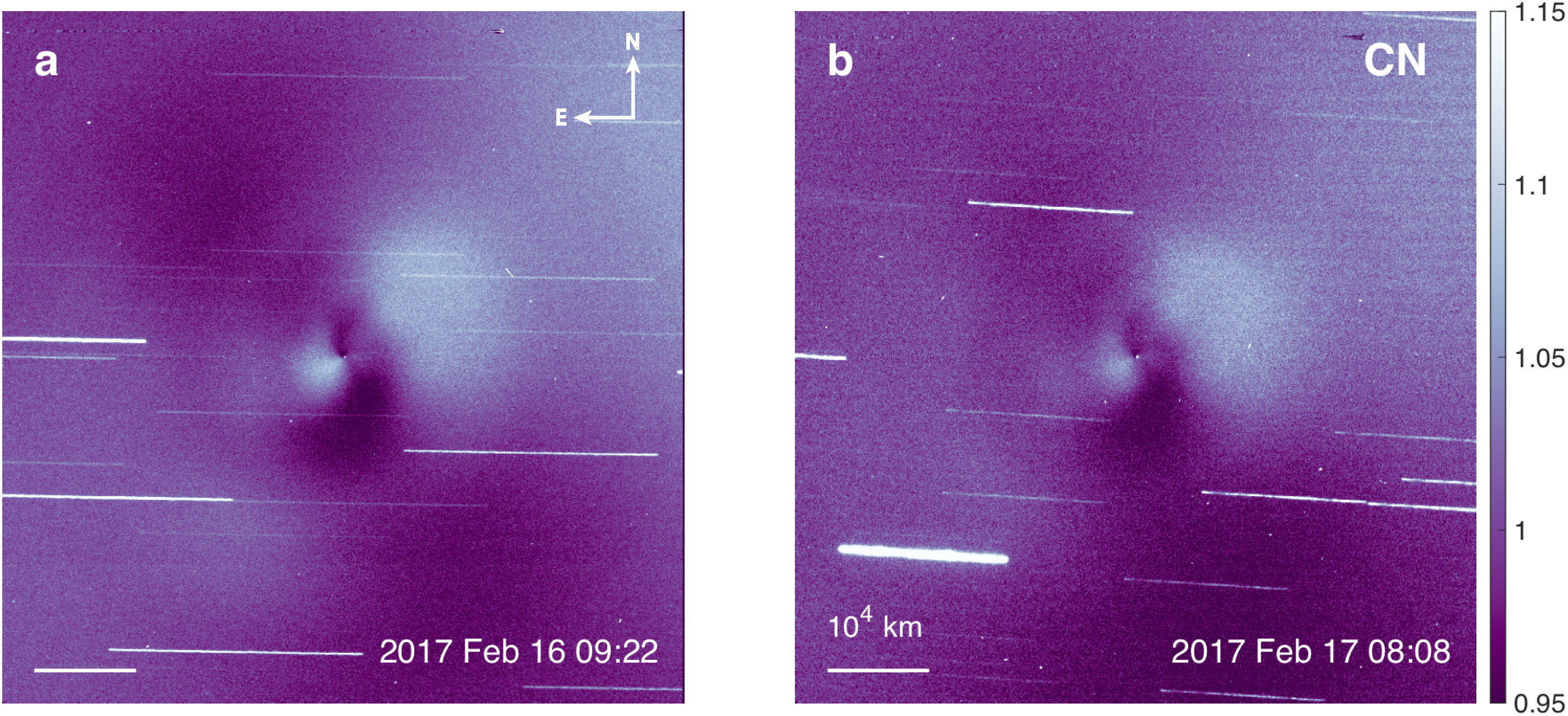}
\caption{Side-by-side comparison of images from 2017 February 16 and 17 taken 22.8 h apart.  The February 16 image is a 1200-s exposure; the February 17 image is a 900-s exposure.  The signal-to-noise ratio is lower in the February 17 image than the February 16 image owing to non-photometric conditions and the shorter exposure time.}
\label{fig:sidebyside}
\end{center}
\end{figure*}

\begin{figure*}[htbp]
\begin{center}
\includegraphics[width=2.1\columnwidth]{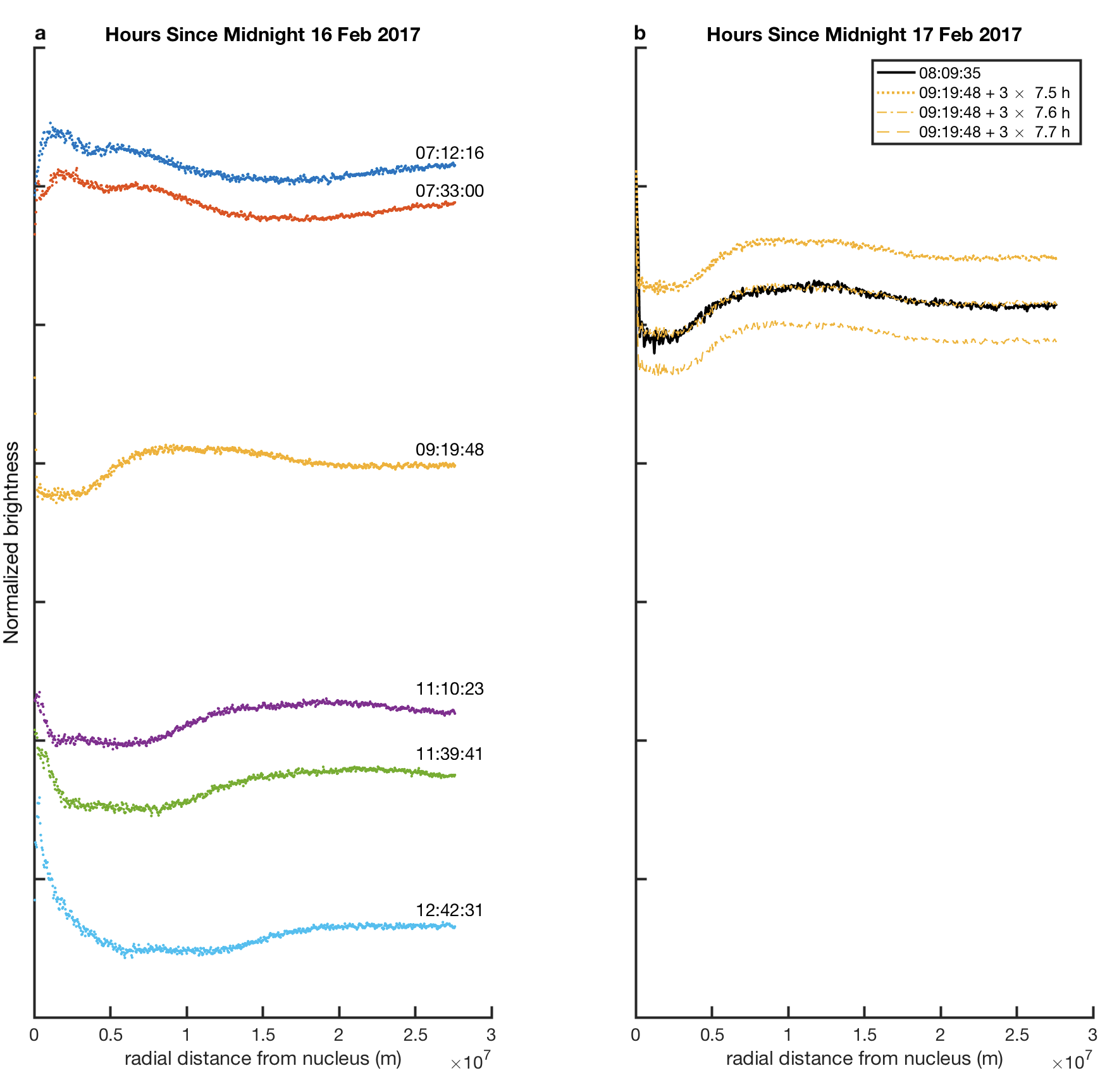}
\caption{Radial profiles of the repeating CN feature in the northwest quadrant of the images taken on February 16 (a) and 17 (b).  Profiles are created by averaging signal over an azimuthal arc of 5$^{\circ}$ centered on the nucleus to reduce signal from star trails and other noise.  The sole profile measured on February 17 (black; b) matches up with a profile taken the previous night (yellow), approximately three rotation periods of $7.6\pm0.1$ hours later.  There is no shifting of the 09:19 February 16 profiles superimposed over the February 17 profile (b).  Vertical separation along the y-axis between profiles is exaggerated to emphasize details in both plots.}
\label{fig:rot_period}
\end{center}
\end{figure*}

The duration of the observations on the first night (5.8 h) is less than one full nucleus rotation (7.6$\pm$0.5 h, E.S.~Howell, personal communication).  We fortuitously have CN images from the first and second night taken 22.8 h or approximately three rotation cycles apart (Figures \ref{fig:sidebyside}, \ref{fig:rot_period}).  The repeating feature is visible in the same position in both images.  This periodicity is consistent with a repeating event releasing material from the nucleus into the inner coma.

\begin{figure*}[htbp]
\begin{center}
\includegraphics[width=2.1\columnwidth]{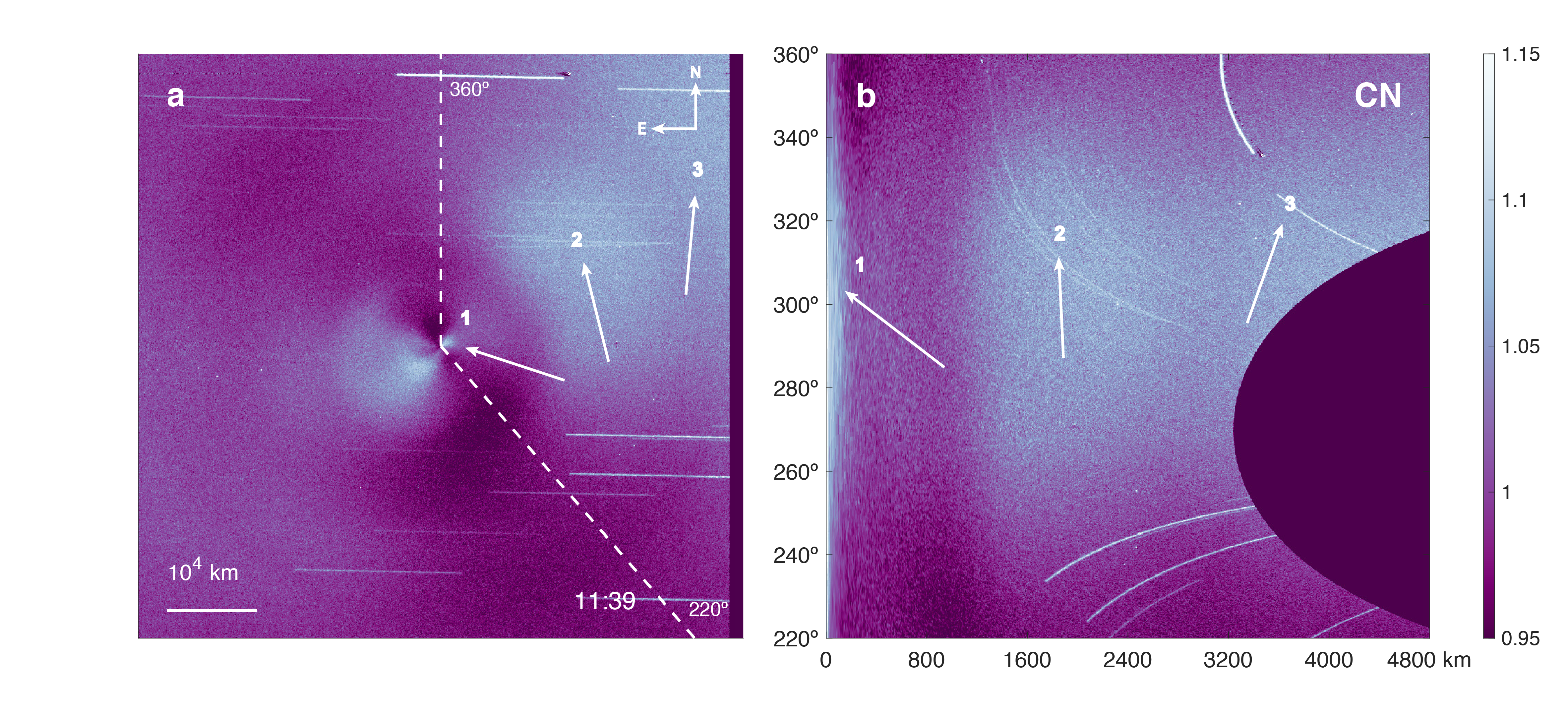}
\caption{Enhanced image of 45P inner coma taken at 11:39 UT on 2017 February 16, with the angular region from 220--360$^{\circ}$ position angle highlighted (a) and a portion of the image unwrapped and projected into position angle-radial distance space from 220--360$^{\circ}$ (b).  Three separate CN emission events are labeled; event 1 is closest to the nucleus.}
\label{fig:unwrap}
\end{center}
\end{figure*}

We measured radial profiles (Figure \ref{fig:rot_period}) of the material moving to the northwest of the nucleus, averaging over a wedge of 5$^{\circ}$ to reduce signal from star trails.  The radial profiles show a two-peaked mass of material moving away from the nucleus into the coma.  Subsequent observations show the peaks broadening and dissipating as the cloud of material retreats from the nucleus.  Despite coverage gaps in the imaging sequence, a second emission of CN gas becomes visible and leads to the bright signal visible in raw images (Figure \ref{fig:raw}).

Unwrapping the enhanced images into projected position angle-radial distance space allows for improved tracking of material moving radially and thus visibility of the three discrete clouds of material expanding away from the nucleus (Figure \ref{fig:unwrap}).  Three discrete clouds of material are visible moving away from the nucleus.
Because of the discontinuities in emission intensity visible in the cloud radial profiles, it is possible to measure the expansion velocity of retreating CN gas.  We measured the center of the expanding cloud moving at a projected velocity of $\sim$0.8 km s$^{-1}$, approximately the velocity of gas expansion at 1 au \citep{Budzien1994}.  The sky plane-projected velocity would be nearly the same as the actual outflow velocity for broad features \citep[e.g.,][]{Samarasinha2000}.  The cloud itself has an expansion velocity of $\sim$0.5 km s$^{-1}$.  

\subsection{C$_2$ and OH features}
\label{sec:otherwavelengths}
\begin{figure*}
\begin{center}
\includegraphics[width=2.1\columnwidth]{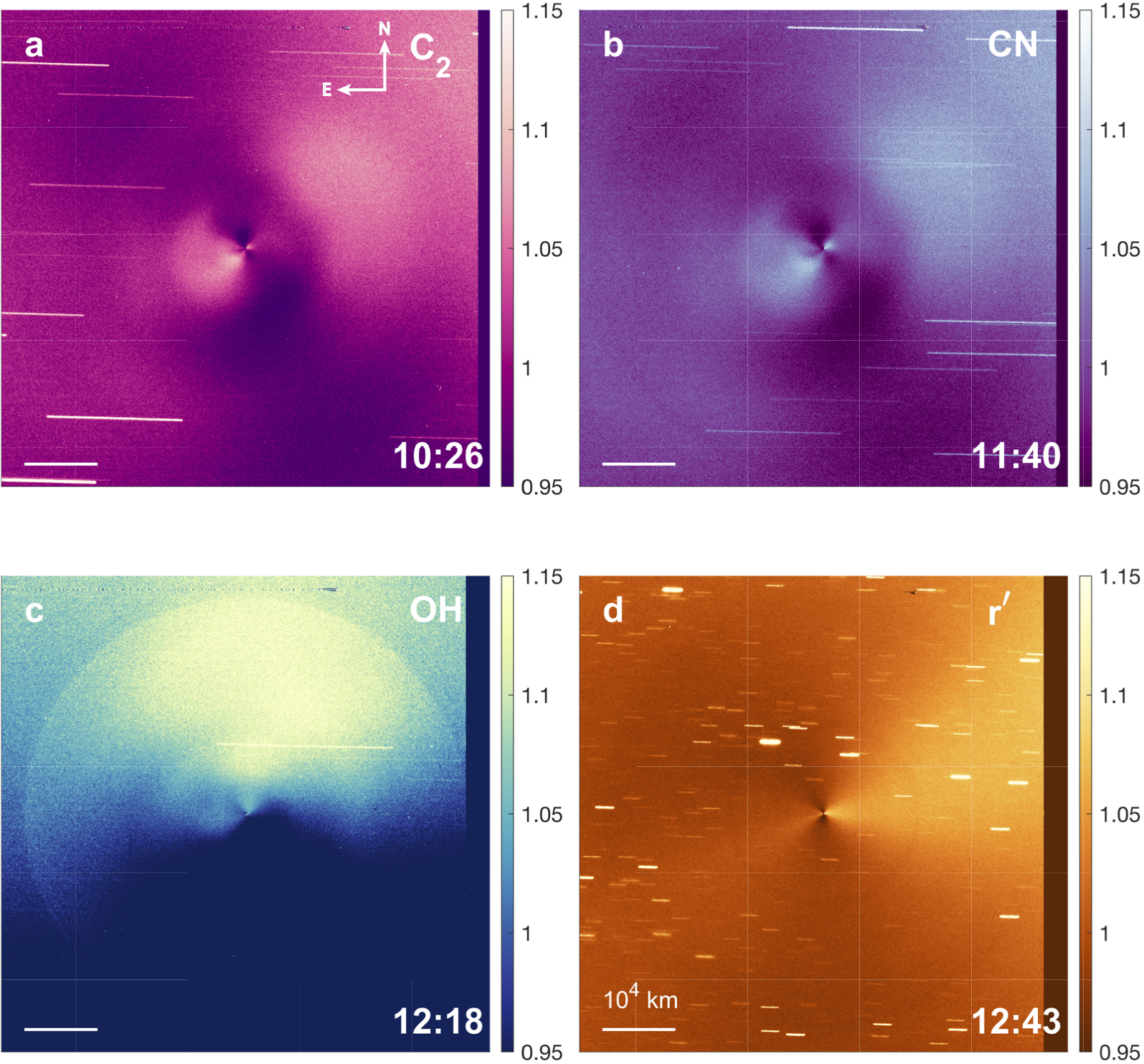}
\caption{Four enhanced images of 45P in C$_2$ (a), CN (b), OH (c), and dust (d) filters at the same distance scales, all taken on 2017 February 16 UT. \imageinfo\ The C$_2$ image  (a) shows two features, a strong one in the northwest, and a fainter one in the southeast.  The CN image (b) shows an southeast jet and a discontinuous northwest jet. The OH image (c) is dominated by a plume to the north ($\sim350^{\circ}$) not seen in other filters, as well as a faint east-pointing jet.  The dust image (d) is dominated by a bright northwest jet and a dimmer southeast jet, which is not the direction of the dust trail.}
\label{fig:multiwavelength}
\end{center}
\end{figure*}

The expanding cloud appears to be present in the C$_2$ images, as well as the southeast jet structure, though the greater continuum background in the C$_2$ images makes detection by visual inspection more difficult (Figure \ref{fig:multiwavelength}a).  
C$_2$ has a broader spatial distribution than CN, which is consistent with CN being primarily a daughter (second-generation) product, whereas C$_2$ is primarily a granddaughter (third-generation) product and thus has more opportunity to retreat from the nucleus and increase its spread.  

The bright compact feature that we observed in an un-enhanced CN image (Figure \ref{fig:raw}) does not appear in the un-enhanced C$_2$ images, making it difficult to constrain the mass of C$_2$ released.  \citet{Moulane2018} measured the C$_2$/CN production rate as $0.04 \pm 0.003$; we therefore estimate that C$_2$ adds 5--10\% of the mass of CN to the total amount of mass ejected ($\sim$\ctwomass{} kg).

Two jet structures are visible in the OH image (Figure \ref{fig:multiwavelength}c): a strong jet pointing approximately 350$^{\circ}$, and a fainter jet pointing in the southeast direction ($\sim100^{\circ}$).  The latter jet corresponds in position angle to the southeast jet in the CN images.  There there is no evidence for an OH cloud in the northwest direction that would correspond to jets seen in either the carbon or dust filters.

Although the jet features visible in the r$^{\prime}$, BC, C$_2$, and CN images are in similar positions to the southeast and northwest of the nucleus, there is only a faint structure at a similar position angle for the OH jet.  Indeed, the stronger OH jet pointing N and carbon species emission to the E are offset by 110$^{\circ}$ as projected into the image plane.  

The curvature of the southeast jet seen in C$_2$, CN, and particularly OH suggests that the positive angular momentum vector is toward the viewer in the northwest quadrant of these images, as OH is constantly produced while in sunlight.  

The northwest dust jet appears continuous, without evidence of the discontinuous clouds seen in CN (Figure \ref{fig:multiwavelength}).  The larger outflow velocity of CN compared to dust could obscure potential discontinuous emission of dust. 

\subsection{Dust production}
\label{sec:dust}
Two jets are visible in the filters representing dust (\rprime\ and BC): one to the southeast ($120^{\circ}$) and a second, stronger jet toward the northwest ($280^{\circ}$; Figure \ref{fig:multiwavelength}).  The dust, CN, and C$_2$ jets are oriented in the same directions.

We, however, do \emph{not} see evidence for a large fragment or other repeating feature in dust wavelengths.  Thus, the process producing the repeating CN feature is creating material only observable in the CN wavelengths.  We therefore conclude that this repeating feature consists primarily of CN. It may contain some C$_2$ gas, despite images of the latter lacking evidence for the discontinuities seen in the CN wavelengths.

As the BC images show the same features as the \rprime\ images, we limit the analysis presented here to \rprime-filter images, which have more temporal coverage and better signal-to-noise ratios with shorter exposure times. Although \rprime\ is a broadband filter and has some gas contamination,the flux in the \rprime images is dominated by scattered dust, not by gas,  unless the comet is very dust-poor.  
The narrowband BC filter was designed to measure the continuum (\textit{i.e.} by the dust). The use of \rprime\ as a proxy for dust has been successfully established for other comets \citep[\textit{e.g.}][]{Mueller2013,Knight2021}.
Images taken in \rprime\ require shorter integration times than BC to achieve high SNR, and also have shorter star trails than BC.  Otherwise, the images are indistinguishable.

\section{Discussion} \label{sec:discussion}

\subsection{Possible causes of the repeating jet feature}
The repeating CN feature to the northwest represents a short-duration release of a small quantity of material.  
If an icy parent molecule, including but not limited to HCN, is emitted from the nucleus that photodissociates to CN, the mass of the parent molecules could be substantially larger, depending on the time of parent molecule release.  
Considering that an unseen parent such as HCN could be the majority of the mass in this feature, but we cannot detect HCN in our filter set, \cnmass{} kg is the lower limit of the mass required to form this visible repeating cloud feature.  
Regardless of whether the CN is emitted directly from the nucleus in gaseous form, or photodissociates from a icy parent molecule, the rapid conversion of this material from ice to gas results in a high production rate for the periodic emission of CN.  

Given the small quantity of CN and C$_2$ ($\sim$\totalmass{} kg; Section \ref{sec:brightcompact}) required to produce the repeating feature, the mechanisms responsible may be simple.  
Whatever the source for the repeating jet feature seen during these observations, it remains the same during the the time span of the observations, and does not show appreciable depletion during the multiple rotational cycles in which the inner coma was observed.  Thus, a single one-time event on the comet nucleus is unlikely to cause the repeating feature.

\begin{figure}[htbp]
\begin{center}
\includegraphics[width=1.05\columnwidth]{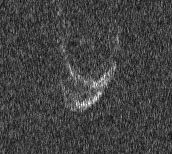}
\caption{A delay-Doppler observation of the 45P comet nucleus taken by the Arecibo Observatory S-band planetary radar system on 2017 February 12 UT.  The comet is 0.65 km in radius.  
Each pixel in the vertical direction represents 60 m. The nucleus is illuminated from the bottom. 
In delay-Doppler projection, distance increases from the bottom of the image, and rotation velocity of the object increases from the center horizontally.
This image shows vertical relief at the $\sim100$-m scale, with steep features deviating from an ellipsoidal nucleus. Credit: Arecibo Observatory/NSF.}
\label{fig:radar}
\end{center}
\end{figure}

The shape of the nucleus of 45P in Arecibo Observatory planetary radar system observations (S-band, 12.6 cm) shows varied near-surface topography at the 100-m scale (Figure \ref{fig:radar}, E.S.~Howell, personal communication).  The existence of cliffs, pits, or crevasses would be consistent with the delay-Doppler data taken of this comet during the same timeframe as the visible-wavelength observations presented here. Exposed ices on nucleus topography would sublimate when exposed to the Sun, potentially producing the repeating northwest-oriented CN feature.

The source region receives enough solar insolation to produce sufficient CN to create this repeating feature. Formation of the feature may occur either by heating a high- to medium-thermal inertia region all day, or by near-instantaneous heating of the source region after being shadowed for the rest of the rotation period.  The all-day scenario requires more topographic relief in the source region, such as pits or cliffs, than instantaneous release of CN-bearing material. 

The $\sim3.31 \times 10^5$ km scale length of CN molecules at 1.05 au \citep{Cochran1985} is larger than the radial distances from the nucleus to the edges of our images ($\sim3 \times10^4$ km; e.g., Figure \ref{fig:cnprogression}).  Either the parent molecule of CN has a short lifetime, or CN originating directly from the nucleus \citep{Hanni2020} could explain the repeating jet and other CN features so close to the optocenters of the images.  

The detection of this repeating feature in CN and C$_2$ but not in other wavelengths may be consistent with heterogeneities in source regions of volatile species, as has been observed by space missions at other comets.  The EPOXI mission at comet {103P/Hartley 2} observed different source locations for several volatile species on the nucleus \citep{AHearn2011,DelloRusso2011}.

Although we do not observe a periodic, repetitive release of dust from 45P, we do see a fan of dust extending in the direction of the northwest feature seen in CN and C$_2$ wavelengths.  This implies a possible relationship between the periodic CN emission and the constant jet of dust.  This could be consistent with larger dust grains drifting at a slower velocity than the smaller CN and C$_2$ molecules: diurnal variation would effectively ``smear'' the dust.  

For {Hartley 2}, \citet{Knight2013} detected similar source regions for CN and other carbon species, as well as OH and NH, but concluded that the latter were likely derived from smaller grains affected by radiation pressure.  \citet{Moretto2017} observed changes at {9P/Tempel 1} in CO$_2$ abundance while H$_2$O remained constant, as well as a correlation between the CO$_2$ and dust distributions; a similar mechanism could be at work for 45P. 

The Rosetta mission to comet 67P/Churyumov--Gerasimenko observed that gas emission and volatile source regions are distributed non-uniformly over the nucleus surface \citep{Hassig2015,Bockelee2015, Migliorini2016,Fink2016,Fougere2016}.  Further, Rosetta observations showed that jet formation corresponds to active pits \citep{Vincent2015} 
and dust and gas emission varies with temperature conditions (solar illumination) \citep{AliLagoa2015,Lara2015}.  
The spacecraft also detected gas-only outbursts from 67P \citep{Feldman2016}.  

Other comets exhibit structures in their gas features, particularly CN jets that can form Archimedean spirals, including 1P/Halley \citep{Klavetter1992}, Hale-Bopp \citep{Mueller1997}, and 103P/Hartley 2 \citep{Samarasinha2011}.  At 45P, the motion of the cloud away from the 45P nucleus, and its expansion, do not resemble an Archimedean spiral.   There is precedent for discontinuities in material emissions into the inner coma: observing on C/1996 B2 (Hyakutake),  \citet{Harris1996} reported a ``detached sunward facing plume.''  Several jet features could have formed into spirals or repeating features; the orientation was such that they would have been difficult to observe from Earth.   Further, \citet{Wierzchos2020} fond that outbursts of CO do not always correlate with dust.

\section{Conclusions \& Future Work} \label{sec:conclusions}
We analyzed 19 images of comet 45P taken 2017 February 16--17 near perigee in the CN, C$_2$, OH, and BC Hale-Bopp campaign narrowband filters, as well as r$^{\prime}$, in conjunction with a multi-wavelength observing campaign. We found the following:
\begin{itemize} 
    \item A bright compact spot adjacent to the nucleus in the CN filter shows comet-like behavior, indicating a rapid release of a CN parent product material, with a total mass of $\sim$\totalmass{ kg.}
    \item This small amount of material could be released from changing solar illumination of surface topology on the nucleus or via another simple process: due to the low mass, a complex mechanism need not be responsible for the bright compact spot.
    \item A high-contrast, repeating CN feature associated with the bright compact spot retreats away from the nucleus at $\sim0.8$ km s$^{-1}$, corresponding to the gas expansion velocity.
    \item The time interval between repetition of the feature is similar to the radar-derived rotation period ($7.6 \pm 0.1$ h) for 45P.
    \item Different position angles for jets from different gas and dust species implies source region heterogeneity on the nucleus for these species.
\end{itemize}

The repeating CN feature may represent a process common on JFCs as they approach the Sun, yet typically obscured from ground-based observations owing to outer coma species obscuring gases such as CN, which may originate directly from the nucleus.  Owing to the low spatial resolutions of ground-based observations, investigation of how common repeating features such as this one are may prove challenging. 

Comets 41P/Tuttle-Giacobini-Kres\'{ak (41P) and 46P/Wirtanen (46P), two other small JFCs, passed within 0.14 au of Earth in 2016--2019 and were observed as part of the coordinated, multi-wavelength campaign in which we observed 45P.  
Image reduction, enhancement, and analysis of narrowband visible-wavelength observations of 41P and 46P are ongoing.
Nevertheless,} we encourage coordinated, multi-wavelength campaigns conducted with medium-sized telescopes for future close approaches of Jupiter-family or other comets to explore whether features like the one observed at 45P are characteristic of comets as they approach perihelion, or whether this behavior is unique to 45P.

\begin{acknowledgments}
We are honored to be permitted to conduct astronomical research on Iolkam Du'ag (Kitt Peak), a mountain with particular significance to the Tohono O'odham Nation in Southwestern Arizona. 

We are grateful to the Steward Observatory TAC for granting us 60+ total nights of observation time in the spring 2017 semester as part of the larger 41P/45P/46P campaign.  We appreciate the expertise of the Kitt Peak National Observatory and Steward Observatory staff and faculty, in particular Joseph Hoscheidt, Melanie Waidanz, Scott Swindell, and Elizabeth ``Betsy'' Green.  We also thank the maintenance/custodial staff of the telescope buildings and dormitories.

The authors thank the Arecibo Observatory staff and scientists who helped take the delay-Doppler data, and express our condolences to the greater Arecibo community on the loss of the extraordinary 305-m William E.~Gordon Telescope in December 2020.  % The authors do not thank the forces that underfunded and led to years of neglect at Arecibo Observatory, resulting in its downfall.

AS thanks Ross A.~Beyer, Cl\'{e}ment Feller, Sierra N.~Ferguson, Hamish C.F.C.~Hay, Margaret E.~Landis, Tod R.~Lauer, Raphael Marschall, Laura C.~Mayorga, Henry Ngo, Timothy E.~Pickering, Christopher W.~Porter, and Maria E.~Steinrueck for useful science conversations and assistance with data reduction.  C.~Wolner provided editing assistance.

This research has made use of NASA's Astrophysics Data System Bibliographic Services, the AstroBetter blog and wiki, the JPL Horizons On-Line Ephemeris System, and Astropy, a community-developed core Python package for astronomy.

% The authors thank the medical professionals and scientists who created the mRNA vaccine, essential workers, medial professionals, first responders, and everyone who ensured we were able to safely work at home throughout the pandemic.
Observations were supported by NASA Solar System Observations grant NNX16AG70G; this analysis was supported in part by NASA FINESST award 80NSSC20K1374.

\end{acknowledgments}

\vspace{5mm}
\facilities{Bok 90'' (90Prime), Arecibo Observatory (Planetary Radar System)}

\software{astropy \citep{Astropy2013,Astropy2018}; sbpy \citep{Mommert2019}}


\begin{thebibliography}{}
\expandafter\ifx\csname natexlab\endcsname\relax\def\natexlab#1{#1}\fi
\providecommand{\url}[1]{\href{#1}{#1}}
\providecommand{\dodoi}[1]{doi:~\href{http://doi.org/#1}{\nolinkurl{#1}}}
\providecommand{\doeprint}[1]{\href{http://ascl.net/#1}{\nolinkurl{http://ascl.net/#1}}}
\providecommand{\doarXiv}[1]{\href{https://arxiv.org/abs/#1}{\nolinkurl{https://arxiv.org/abs/#1}}}

\bibitem[{A{\textquoteright}Hearn {et~al.}(2011)A{\textquoteright}Hearn,
  Belton, Delamere, Feaga, Hampton, Kissel, Klaasen, McFadden, Meech, Melosh,
  Schultz, Sunshine, Thomas, Veverka, Wellnitz, Yeomans, Besse, Bodewits,
  Bowling, Carcich, Collins, Farnham, Groussin, Hermalyn, Kelley, Kelley, Li,
  Lindler, Lisse, McLaughlin, Merlin, Protopapa, Richardson, \&
  Williams}]{AHearn2011}
A{\textquoteright}Hearn, M.~F., Belton, M. J.~S., Delamere, W.~A., {et~al.}
  2011, Science, 332, 1396, \dodoi{10.1126/science.1204054}

\bibitem[{Al\'{i}-Lagoa {et~al.}(2015)Al\'{i}-Lagoa, Delbo, \&
  Libourel}]{AliLagoa2015}
Al\'{i}-Lagoa, V., Delbo, M., \& Libourel, G. 2015, The Astrophysical Journal
  Letters, 810, L22

\bibitem[{{Astropy Collaboration} {et~al.}(2013){Astropy Collaboration},
  {Robitaille}, {Tollerud}, {Greenfield}, {Droettboom}, {Bray}, {Aldcroft},
  {Davis}, {Ginsburg}, {Price-Whelan}, {Kerzendorf}, {Conley}, {Crighton},
  {Barbary}, {Muna}, {Ferguson}, {Grollier}, {Parikh}, {Nair}, {Unther},
  {Deil}, {Woillez}, {Conseil}, {Kramer}, {Turner}, {Singer}, {Fox}, {Weaver},
  {Zabalza}, {Edwards}, {Azalee Bostroem}, {Burke}, {Casey}, {Crawford},
  {Dencheva}, {Ely}, {Jenness}, {Labrie}, {Lim}, {Pierfederici}, {Pontzen},
  {Ptak}, {Refsdal}, {Servillat}, \& {Streicher}}]{Astropy2013}
{Astropy Collaboration}, {Robitaille}, T.~P., {Tollerud}, E.~J., {et~al.} 2013,
  \aap, 558, A33, \dodoi{10.1051/0004-6361/201322068}

\bibitem[{{Astropy Collaboration} {et~al.}(2018){Astropy Collaboration},
  {Price-Whelan}, {Sip{\H{o}}cz}, {G{\"u}nther}, {Lim}, {Crawford}, {Conseil},
  {Shupe}, {Craig}, {Dencheva}, {Ginsburg}, {VanderPlas}, {Bradley},
  {P{\'e}rez-Su{\'a}rez}, {de Val-Borro}, {Aldcroft}, {Cruz}, {Robitaille},
  {Tollerud}, {Ardelean}, {Babej}, {Bach}, {Bachetti}, {Bakanov}, {Bamford},
  {Barentsen}, {Barmby}, {Baumbach}, {Berry}, {Biscani}, {Boquien}, {Bostroem},
  {Bouma}, {Brammer}, {Bray}, {Breytenbach}, {Buddelmeijer}, {Burke},
  {Calderone}, {Cano Rodr{\'\i}guez}, {Cara}, {Cardoso}, {Cheedella}, {Copin},
  {Corrales}, {Crichton}, {D'Avella}, {Deil}, {Depagne}, {Dietrich}, {Donath},
  {Droettboom}, {Earl}, {Erben}, {Fabbro}, {Ferreira}, {Finethy}, {Fox},
  {Garrison}, {Gibbons}, {Goldstein}, {Gommers}, {Greco}, {Greenfield},
  {Groener}, {Grollier}, {Hagen}, {Hirst}, {Homeier}, {Horton}, {Hosseinzadeh},
  {Hu}, {Hunkeler}, {Ivezi{\'c}}, {Jain}, {Jenness}, {Kanarek}, {Kendrew},
  {Kern}, {Kerzendorf}, {Khvalko}, {King}, {Kirkby}, {Kulkarni}, {Kumar},
  {Lee}, {Lenz}, {Littlefair}, {Ma}, {Macleod}, {Mastropietro}, {McCully},
  {Montagnac}, {Morris}, {Mueller}, {Mumford}, {Muna}, {Murphy}, {Nelson},
  {Nguyen}, {Ninan}, {N{\"o}the}, {Ogaz}, {Oh}, {Parejko}, {Parley}, {Pascual},
  {Patil}, {Patil}, {Plunkett}, {Prochaska}, {Rastogi}, {Reddy Janga},
  {Sabater}, {Sakurikar}, {Seifert}, {Sherbert}, {Sherwood-Taylor}, {Shih},
  {Sick}, {Silbiger}, {Singanamalla}, {Singer}, {Sladen}, {Sooley},
  {Sornarajah}, {Streicher}, {Teuben}, {Thomas}, {Tremblay}, {Turner},
  {Terr{\'o}n}, {van Kerkwijk}, {de la Vega}, {Watkins}, {Weaver}, {Whitmore},
  {Woillez}, {Zabalza}, \& {Astropy Contributors}}]{Astropy2018}
{Astropy Collaboration}, {Price-Whelan}, A.~M., {Sip{\H{o}}cz}, B.~M., {et~al.}
  2018, \aj, 156, 123, \dodoi{10.3847/1538-3881/aabc4f}

\bibitem[{Bockel{\'e}e-Morvan {et~al.}(2015)Bockel{\'e}e-Morvan, Debout, Erard,
  Leyrat, Capaccioni, Filacchione, Fougere, Drossart, Arnold, Combi,
  {et~al.}}]{Bockelee2015}
Bockel{\'e}e-Morvan, D., Debout, V., Erard, S., {et~al.} 2015, Astronomy \&
  Astrophysics, 583, A6

\bibitem[{{Budzien} {et~al.}(1994){Budzien}, {Festou}, \&
  {Feldman}}]{Budzien1994}
{Budzien}, S.~A., {Festou}, M.~C., \& {Feldman}, P.~D. 1994, \icarus, 107, 164,
  \dodoi{10.1006/icar.1994.1014}

\bibitem[{{Cochran}(1985)}]{Cochran1985}
{Cochran}, A.~L. 1985, \aj, 90, 2609, \dodoi{10.1086/113966}

\bibitem[{{Dello Russo} {et~al.}(2011){Dello Russo}, {Vervack}, {Lisse},
  {Weaver}, {Kawakita}, {Kobayashi}, {Cochran}, {Harris}, {McKay}, \&
  {Biver}}]{DelloRusso2011}
{Dello Russo}, N., {Vervack}, R.~J., J., {Lisse}, C.~M., {et~al.} 2011, \apj,
  734, L8, \dodoi{10.1088/2041-8205/734/1/L8}

\bibitem[{{Dello Russo} {et~al.}(2020){Dello Russo}, {Kawakita}, {Bonev},
  {Vervack}, {Gibb}, {Shinnaka}, {Roth}, {DiSanti}, \&
  {McKay}}]{DelloRusso2020}
{Dello Russo}, N., {Kawakita}, H., {Bonev}, B.~P., {et~al.} 2020, \icarus, 335,
  113411, \dodoi{10.1016/j.icarus.2019.113411}

\bibitem[{Di~Sisto \& Brunini(2007)}]{DiSisto2007}
Di~Sisto, R.~P., \& Brunini, A. 2007, Icarus, 190, 224,
  \dodoi{https://doi.org/10.1016/j.icarus.2007.02.012}

\bibitem[{DiSanti {et~al.}(2017)DiSanti, Bonev, Russo, Vervack, Gibb, Roth,
  McKay, Kawakita, Feaga, \& Weaver}]{DiSanti2017}
DiSanti, M.~A., Bonev, B.~P., Russo, N.~D., {et~al.} 2017, The Astronomical
  Journal, 154, 246, \dodoi{10.3847/1538-3881/aa8639}

\bibitem[{{Dones} {et~al.}(2015){Dones}, {Brasser}, {Kaib}, \&
  {Rickman}}]{Dones2015}
{Dones}, L., {Brasser}, R., {Kaib}, N., \& {Rickman}, H. 2015, \ssr, 197, 191,
  \dodoi{10.1007/s11214-015-0223-2}

\bibitem[{{Duncan} {et~al.}(2004){Duncan}, {Levison}, \& {Dones}}]{Duncan2004}
{Duncan}, M., {Levison}, H., \& {Dones}, L. 2004, in Comets II, ed. M.~C.
  {Festou}, H.~U. {Keller}, \& H.~A. {Weaver} ({University of Arizona Press}),
  193

\bibitem[{{Duncan} \& {Levison}(1997)}]{Duncan1997}
{Duncan}, M.~J., \& {Levison}, H.~F. 1997, Science, 276, 1670,
  \dodoi{10.1126/science.276.5319.1670}

\bibitem[{Farnham {et~al.}(2000)Farnham, Schleicher, \& A'Hearn}]{Farnham2000}
Farnham, T.~L., Schleicher, D.~G., \& A'Hearn, M.~F. 2000, Icarus, 147, 180 ,
  \dodoi{http://dx.doi.org/10.1006/icar.2000.6420}

\bibitem[{{Feldman} {et~al.}(2016){Feldman}, {A'Hearn}, {Feaga}, {Bertaux},
  {Noonan}, {Parker}, {Schindhelm}, {Steffl}, {Stern}, \&
  {Weaver}}]{Feldman2016}
{Feldman}, P.~D., {A'Hearn}, M.~F., {Feaga}, L.~M., {et~al.} 2016, \apj, 825,
  L8, \dodoi{10.3847/2041-8205/825/1/L8}

\bibitem[{Fink(2009)}]{Fink2009}
Fink, U. 2009, Icarus, 201, 311

\bibitem[{Fink {et~al.}(2016)Fink, Doose, Rinaldi, Bieler, Capaccioni,
  Bockel{\'e}e-Morvan, Filacchione, Erard, Leyrat, Blecka, Capria, Combi,
  Crovisier, Sanctis, Fougere, Taylor, Migliorini, \& Piccioni}]{Fink2016}
Fink, U., Doose, L., Rinaldi, G., {et~al.} 2016, Icarus, 277, 78,
  \dodoi{https://doi.org/10.1016/j.icarus.2016.04.040}

\bibitem[{Fougere {et~al.}(2016)Fougere, {Altwegg, K.}, {Berthelier, J.-J.},
  {Bieler, A.}, {Bockel{\'e}e-Morvan, D.}, {Calmonte, U.}, {Capaccioni, F.},
  {Combi, M. R.}, {De Keyser, J.}, {Debout, V.}, {Erard, S.}, {Fiethe, B.},
  {Filacchione, G.}, {Fink, U.}, {Fuselier, S. A.}, {Gombosi, T. I.}, {Hansen,
  K. C.}, {H{\"a}ssig, M.}, {Huang, Z.}, {Le Roy, L.}, {Leyrat, C.},
  {Migliorini, A.}, {Piccioni, G.}, {Rinaldi, G.}, {Rubin, M.}, {Shou, Y.},
  {Tenishev, V.}, {Toth, G.}, {Tzou, C.-Y.}, {the VIRTIS team}, \& {the ROSINA
  team}}]{Fougere2016}
Fougere, N., {Altwegg, K.}, {Berthelier, J.-J.}, {et~al.} 2016, Astronomy \&
  Astrophysics, 588, A134, \dodoi{10.1051/0004-6361/201527889}

\bibitem[{{H{\"a}nni} {et~al.}(2020){H{\"a}nni}, {Altwegg}, {Pestoni}, {Rubin},
  {Schroeder}, {Schuhmann}, \& {Wampfler}}]{Hanni2020}
{H{\"a}nni}, N., {Altwegg}, K., {Pestoni}, B., {et~al.} 2020, \mnras, 498,
  2239, \dodoi{10.1093/mnras/staa2387}

\bibitem[{{Harris} {et~al.}(1996){Harris}, {Scherb}, {Combi}, \&
  {Mueller}}]{Harris1996}
{Harris}, W.~M., {Scherb}, F., {Combi}, M.~R., \& {Mueller}, B.~E.~A. 1996, in
  AAS/Division for Planetary Sciences Meeting Abstracts, Vol.~28, AAS/Division
  for Planetary Sciences Meeting Abstracts \#28, 09.03

\bibitem[{{Harris} {et~al.}(2017){Harris}, {Ryan}, {Springmann}, {Mueller},
  {Samarasinha}, {Kikwaya Elou}, {Howell}, {Lejoly}, {Bodnarik}, {Fitzpatrick},
  {Maciel}, {Mitchell}, \& {Watson}}]{Harris2017}
{Harris}, W.~M., {Ryan}, E.~L., {Springmann}, A., {et~al.} 2017, in
  AAS/Division for Planetary Sciences Meeting Abstracts \#49, AAS/Division for
  Planetary Sciences Meeting Abstracts, 305.05

\bibitem[{H{\"a}ssig {et~al.}(2015)H{\"a}ssig, Altwegg, Balsiger, Bar-Nun,
  Berthelier, Bieler, Bochsler, Briois, Calmonte, Combi, {et~al.}}]{Hassig2015}
H{\"a}ssig, M., Altwegg, K., Balsiger, H., {et~al.} 2015, Science, 347, aaa0276

\bibitem[{{Howell} {et~al.}(2018){Howell}, {Belton}, {Samarasinha}, {Mueller},
  {Harris}, {Nolan}, {Taylor}, {Rivera-Valentin}, \&
  {Zambrano-Marin}}]{Howell2018}
{Howell}, E., {Belton}, M.~J., {Samarasinha}, N.~H., {et~al.} 2018, in
  AAS/Division for Planetary Sciences Meeting Abstracts, 106.06

\bibitem[{Klavetter \& A'Hearn(1992)}]{Klavetter1992}
Klavetter, J.~J., \& A'Hearn, M.~F. 1992, Icarus, 95, 73 ,
  \dodoi{https://doi.org/10.1016/0019-1035(92)90192-A}

\bibitem[{Knight \& Schleicher(2013)}]{Knight2013}
Knight, M.~M., \& Schleicher, D.~G. 2013, Icarus, 222, 691

\bibitem[{Knight {et~al.}(2021)Knight, Schleicher, \& Farnham}]{Knight2021}
Knight, M.~M., Schleicher, D.~G., \& Farnham, T.~L. 2021, The Planetary Science
  Journal, 2, 104, \dodoi{10.3847/psj/abef6c}

\bibitem[{Lara {et~al.}(2015)Lara, Lowry, Vincent, Guti{\'e}rrez, Ro{\.z}ek,
  La~Forgia, Oklay, Sierks, Barbieri, Lamy, {et~al.}}]{Lara2015}
Lara, L., Lowry, S., Vincent, J.-B., {et~al.} 2015, Astronomy \& Astrophysics,
  583, A9

\bibitem[{{Lejoly} {et~al.}(2017){Lejoly}, {Howell}, \&
  {Zambrano-Marin}}]{CBET2017}
{Lejoly}, C., {Howell}, E., \& {Zambrano-Marin}, F. 2017, Central Bureau
  Electronic Telegrams, 4357, 1

\bibitem[{Migliorini {et~al.}(2016)Migliorini, Piccioni, Capaccioni,
  Filacchione, Bockel{\'e}e-Morvan, Erard, Leyrat, Combi, Fougere, Crovisier,
  {et~al.}}]{Migliorini2016}
Migliorini, A., Piccioni, G., Capaccioni, F., {et~al.} 2016, Astronomy \&
  Astrophysics, 589, A45

\bibitem[{{Mommert} {et~al.}(2019){Mommert}, {Kelley}, {de Val-Borro}, {Li},
  {Guzman}, {Sipo{\`I}cz}, {D{\`I}urech}, {Granvik}, {Grundy}, {Moskovitz},
  {Penttil{\"a}}, \& {Samarasinha}}]{Mommert2019}
{Mommert}, M., {Kelley}, M. S.~P., {de Val-Borro}, M., {et~al.} 2019, {sbpy:
  Small-body planetary astronomy}.
\newblock \doeprint{1907.014}

\bibitem[{{Moretto} {et~al.}(2017){Moretto}, {Feaga}, \&
  {A'Hearn}}]{Moretto2017}
{Moretto}, M.~J., {Feaga}, L.~M., \& {A'Hearn}, M.~F. 2017, \icarus, 296, 28,
  \dodoi{10.1016/j.icarus.2017.05.013}

\bibitem[{Moulane {et~al.}(2018)Moulane, Jehin, Opitom, Pozuelos, Manfroid,
  Benkhaldoun, Daassou, \& Gillon}]{Moulane2018}
Moulane, Y., Jehin, E., Opitom, C., {et~al.} 2018, A\&A, 619, A156,
  \dodoi{10.1051/0004-6361/201833582}

\bibitem[{Mueller {et~al.}(1997)Mueller, Samarasinha, \& Belton}]{Mueller1997}
Mueller, B.~E., Samarasinha, N.~H., \& Belton, M.~J. 1997, Earth, Moon, and
  Planets, 77, 181, \dodoi{10.1023/A:1006286830412}

\bibitem[{Mueller {et~al.}(2013)Mueller, Samarasinha, Farnham, \&
  A'Hearn}]{Mueller2013}
Mueller, B.~E., Samarasinha, N.~H., Farnham, T.~L., \& A'Hearn, M.~F. 2013,
  Icarus, 222, 799

\bibitem[{Samarasinha(2000)}]{Samarasinha2000}
Samarasinha, N.~H. 2000, The Astrophysical Journal Letters, 529, L107.
\newblock \url{http://stacks.iop.org/1538-4357/529/i=2/a=L107}

\bibitem[{Samarasinha \& Larson(2014)}]{Samarasinha2014}
Samarasinha, N.~H., \& Larson, S.~M. 2014, Icarus, 239, 168

\bibitem[{Samarasinha {et~al.}(2011)Samarasinha, Mueller, A'Hearn, Farnham, \&
  Gersch}]{Samarasinha2011}
Samarasinha, N.~H., Mueller, B. E.~A., A'Hearn, M.~F., Farnham, T.~L., \&
  Gersch, A. 2011, The Astrophysical Journal, 734, L3,
  \dodoi{10.1088/2041-8205/734/1/l3}

\bibitem[{Sarid {et~al.}(2019)Sarid, Volk, Steckloff, Harris, Womack, \&
  Woodney}]{Sarid2019}
Sarid, G., Volk, K., Steckloff, J.~K., {et~al.} 2019, The Astrophysical
  Journal, 883, L25, \dodoi{10.3847/2041-8213/ab3fb3}

\bibitem[{{Schleicher}(2010)}]{Schleicher2010}
{Schleicher}, D.~G. 2010, \aj, 140, 973, \dodoi{10.1088/0004-6256/140/4/973}

\bibitem[{{Schleicher} {et~al.}(2019){Schleicher}, {Knight}, {Eisner}, \&
  {Thirouin}}]{Schleicher2019}
{Schleicher}, D.~G., {Knight}, M.~M., {Eisner}, N.~L., \& {Thirouin}, A. 2019,
  \aj, 157, 108, \dodoi{10.3847/1538-3881/aafbab}

\bibitem[{Steckloff {et~al.}(2020)Steckloff, Sarid, Volk, Kareta, Womack,
  Harris, Woodney, \& Schambeau}]{Steckloff2020}
Steckloff, J.~K., Sarid, G., Volk, K., {et~al.} 2020, The Astrophysical Journal
  Letters, 904, L20, \dodoi{10.3847/2041-8213/abc888}

\bibitem[{{Tiscareno} \& {Malhotra}(2003)}]{Tiscareno2003}
{Tiscareno}, M.~S., \& {Malhotra}, R. 2003, \aj, 126, 3122,
  \dodoi{10.1086/379554}

\bibitem[{Vincent {et~al.}(2015)Vincent, Bodewits, Besse, Sierks, Barbieri,
  Lamy, Rodrigo, Koschny, Rickman, Keller, {et~al.}}]{Vincent2015}
Vincent, J.-B., Bodewits, D., Besse, S., {et~al.} 2015, Nature, 523, 63

\bibitem[{{Wierzchos} \& {Womack}(2020)}]{Wierzchos2020}
{Wierzchos}, K., \& {Womack}, M. 2020, \aj, 159, 136,
  \dodoi{10.3847/1538-3881/ab6e68}

\end{thebibliography}
\end{document}